\documentclass[aps,prx,onecolumn,superscriptaddress,longbibliography]{revtex4-2}

\usepackage{amsmath,amsfonts,bbm}
\usepackage[colorlinks=true,breaklinks=true,linkcolor=magenta,citecolor=magenta,urlcolor=magenta]{hyperref}
\usepackage{comment}
\usepackage{graphicx}
\usepackage{floatrow}
\usepackage{mathtools}
\usepackage{booktabs}
\usepackage{diagbox}
\usepackage{listings}
\usepackage{parskip}
\usepackage{braket}
\usepackage{makecell}
\usepackage{tabularx}
\usepackage{verbatim}
\usepackage{float}
\usepackage{url}
\usepackage{subcaption}
\usepackage{caption}
\usepackage{algorithm}
\usepackage{algpseudocode}
\usepackage[bb=boondox]{mathalfa}
\usepackage{tikz}
\usetikzlibrary{arrows.meta}
\usetikzlibrary{quantikz}
\usetikzlibrary{calc, arrows.meta, shapes.geometric}
\usepackage{enumitem} 
\usepackage{todonotes}

\newtheorem{definition}{Definition}

\begin{document}

\title{Evaluating Sample-Based Krylov Quantum Diagonalization for Heisenberg Models with Applications to Materials Science}

\author{Roman Firt}
\email{roman.firt@volkswagen.de}
\affiliation{Volkswagen AG, Berliner Ring 2, Wolfsburg 38440, Germany}

\author{Neel Misciasci}
\affiliation{Volkswagen AG, Berliner Ring 2, Wolfsburg 38440, Germany}
\affiliation{CIT School, Technical University of Munich, Germany}

\author{Jonathan E. Mueller}
\affiliation{Volkswagen AG, Berliner Ring 2, Wolfsburg 38440, Germany}

\author{Triet Friedhoff}
\affiliation{IBM Quantum, IBM Chicago Office, Chicago, IL 60606, USA}

\author{Chinonso Onah}
\affiliation{Volkswagen AG, Berliner Ring 2, Wolfsburg 38440, Germany}
\affiliation{Department of Physics, RWTH Aachen, Germany}

\author{Aaron Schulze}
\affiliation{Volkswagen AG, Berliner Ring 2, Wolfsburg 38440, Germany}
\affiliation{University of Ulm, Department of Physics, Albert-Einstein-Allee 11, Ulm 89081, Germany}

\author{Sarah Mostame}
\email{sarah.mostame@ibm.com}
\affiliation{IBM Quantum, IBM T.J. Watson Research Center, Yorktown Heights, New York 10598, USA}

\date{\today}

\begin{abstract}
We evaluate the Sample-based Krylov Quantum Diagonalization (SKQD) algorithm on one- and two-dimensional Heisenberg models, including strongly correlated regimes in which the ground state is dense.
Using problem-informed initial states and magnetization-sector sweeps, SKQD accurately reproduces ground-state energies and field-dependent magnetization across a range of anisotropies. 
Benchmarks against DMRG and exact diagonalization show consistent qualitative agreement, with accuracy improving systematically in more anisotropic regimes. 
We further demonstrate SKQD on quantum hardware by implementing 18- and 30-qubit Heisenberg chains, obtaining magnetization curves that match theoretical expectations. 
Simulations on small 2D square-lattice systems further demonstrate that the method applies effectively beyond 1D geometries. 
%
\end{abstract}

\maketitle

\section{Introduction and background}

The Heisenberg model is a cornerstone of quantum magnetism, capturing the essential physics of interacting localized spins. It serves as a fundamental framework for understanding quantum many-body phenomena such as entanglement, quantum phase transitions, and emergent collective behavior in both low- and higher-dimensional systems. In particular, the spin-$\tfrac{1}{2}$ Heisenberg Hamiltonian provides a minimal yet non-trivial setting for studying strongly correlated quantum matter that can be probed analytically, simulated numerically, and realized experimentally on both classical and quantum platforms. 
    Beyond its historical importance, the quantum Heisenberg model has become a central benchmark for the development of quantum simulation protocols on quantum hardware \cite{PhysRevResearch.5.013183,Keenan_2023,Elliot,yoshioka2025krylov,kumaran2025}. Its tunable ground-state structure, ranging from critical entangled states in the gapless Luttinger-liquid regime to trivially polarized states under strong magnetic fields, offers a unique testbed for quantum algorithms. Variations in anisotropy and external fields control the transition between sparse and non-sparse representations of the Hamiltonian, making the model a suitable playground for evaluating algorithmic robustness and scalability on quantum devices.

The generic Heisenberg Hamiltonian for a lattice of spin-$\tfrac{1}{2}$  particles is given by
  \begin{equation}
    H = J \, \sum_{\langle i,j \rangle} 
    \left( S_i^x S_j^x + S_i^y S_j^y + \Delta \, S_i^z S_j^z \right) 
    - \sum_i \vec{h} \cdot \vec{S}_i 
    \label{eq:genericHamiltonian}
  \end{equation}
where \mbox{$S^\alpha (\alpha=x,y,z)$} are Pauli matrices, ${\langle i,j \rangle} $ denotes pairs of neighboring spins, $\vec{h}$ is an external magnetic field, with $J$ being the exchange coupling constant; $J>0$ represents the antiferromagnetic (AFM) coupling, while $J<0$ describes the ferromagnetic (FM) coupling. $\Delta$ is the anisotropy parameter; $\Delta = 1$ describes the isotropic Heisenberg model ($XXX$ model) while $\Delta \ne 1 $ represents the anisotropic model ($XXZ$ model). Exemplary values of the exchange coupling $J$ and the anisotropy parameter $\Delta$ for real spin-$\tfrac{1}{2}$ quasi-1D quantum magnets are listed in table (\ref{tab:realmaterialsvalues}). Values reaching from easy-plane $XX$-anisotropy with $\Delta<1$ to isotropic materials with $\Delta=1$ up to Ising-type anisotropy with $\Delta \gg 1$, which shows that the full range of anisotropies in the XXZ model is significant for real materials.

\begin{table}[H]
\centering
\begin{tabular}{lccc}
\hline
\textbf{Material} & $J/\mathrm{meV}$ & $\Delta$ & \textbf{Reference} \\
\hline
Cs$_2$CoCl$_4$ & 0.23 & 0.25 &
\cite{PhysRevLett.127.037201} \\
CuPzN & 0.91 & 1.00 &
\cite{PhysRevB.59.1008} \\
KCuF$_3$ & 33.5 & 1.00 &
\cite{PhysRevLett.111.137205} \\
BaCo$_2$V$_2$O$_8$ & 3.05 & 1.90 &
\cite{PhysRevLett.123.027204} \\
SrCo$_2$V$_2$O$_8$ & 3.7 & 2.10 &
\cite{PhysRevLett.123.067203} \\
CsCoBr$_3$ & 1.25 & 6.25 &
\cite{WPLehmann_1981} \\
CsCoCl$_3$ & 0.595 & 10.42 &
\cite{WPLehmann_1981} \\
\hline
\end{tabular}
\caption{Reported values of the exchange coupling $J$ and
 anisotropy parameter $\Delta$ for selected spin-$\tfrac{1}{2}$ quasi-1D quantum magnets that can be described with the $XXZ$-Hamiltonian from equation (\ref{eq:genericHamiltonian}). To show comparable results, $\Delta$ was derived from anisotropy values if other parameters were reported.}
 \label{tab:realmaterialsvalues}
\end{table}

A magnetic field applied along the $z$-axis adds a Zeeman term
$ H_{\text{z}} = -h_z \sum_i S_i^z \, , $
which preserves the $U(1)$ symmetry associated with $S_{tot}^z$, but breaks the $SU(2)$ symmetry in the isotropic case. 
It shifts the energy of spin-up vs spin-down states and can polarize the system.
In the strong-field limit, the ground state becomes fully polarized, leading to a trivial product state. 
At weaker fields, the ground state remains highly entangled, but becomes partially polarized.
In the 1D Heisenberg model in Zeeman external field, the system exhibits a quantum phase transition as magnetization increases with field strength.
A magnetic field applied perpendicular to the quantization axis (e.g., in the $x$-direction) introduces a transverse term
$ H_{\text{tr}} = -h_x \sum_i S_i^x \, .$
This term does not commute with $S_z$, leading to quantum fluctuations that can destroy magnetic order.  
In the anisotropic $XXZ$ and Ising limits, transverse fields can induce quantum phase transitions, such as from a Néel-ordered phase to a paramagnet.  
The transverse-field Ising model is a particularly well-studied limiting case with rich critical behavior in 1D and 2D.
The magnetization of a quantum spin system quantifies the degree of alignment of spins with respect to a given direction, and it serves as a key observable to characterize magnetic order and phase transitions.

Despite extensive classical studies of these systems, numerical simulation of strongly correlated and non-sparse Hamiltonians remains computationally demanding. For example, the variational quantum simulation of the 1D Heisenberg model reported in Ref.~\cite{PhysRevResearch.5.013183} achieved notable accuracy but relied on variational ansatze tailored for relatively sparse representations. 

In this work, we extend this line of research by evaluating the recently developed \textit{Sample-based Krylov Quantum Diagonalization} (SKQD)~\cite{yu2025sample} algorithm on the Heisenberg model. The SKQD approach belongs to the emerging class of sample-based quantum diagonalization methods, designed for efficient implementation on quantum-centric supercomputing architectures~\cite{ALEXEEV2024666}. Although the Heisenberg model does not inherently satisfy the sparsity requirement assumed in the formal efficiency analysis of SKQD, we develop a systematic strategy for adapting the algorithm to strongly correlated Hamiltonians with a dense ground-state structure.
Remarkably, this enables reliable estimation of observables such as magnetization across multiple regimes of the phase diagram, including both isotropic and anisotropic interactions and in the presence of external magnetic fields. 
Figure~\ref{fig:sparse} provides a simple yet informative indicator of how concentrated the ground-state probability distribution is, and therefore how challenging the state may be to approximate or sample in algorithms that rely on basis-state sparsity.
Our results suggest that SKQD can be applied beyond 
the sparsity assumptions underlying its theoretical efficiency analysis,
without an immediate loss of practical performance. 
This work thus connects algorithmic innovation with physically rich test systems, bridging current quantum computational techniques and foundational models of quantum magnetism.

\section{method}
Our study focuses on the Sample-based Krylov Quantum Diagonalization (SKQD) algorithm~\cite{yu2025sample}, a recent advance in hybrid quantum–classical computation for spectral estimation of quantum many-body systems. SKQD determines the low-energy spectrum by constructing and diagonalizing an effective Hamiltonian in a Krylov subspace, where all matrix elements are obtained directly from sampled quantum measurements rather than explicit state vectors. This strategy enables scalable and noise-resilient spectral estimation that circumvents the exponential overhead associated with full-state tomography.
Developed to mitigate the limitations of current quantum hardware, SKQD unifies key elements from two complementary approaches: Krylov Quantum Diagonalization (KQD)~\cite{yoshioka2025krylov} and Sample-based Quantum Diagonalization methods (SQD)~\cite{QSCI2023,robledo2025chemistry}. KQD constructs a Krylov subspace via short-time unitary evolution followed by classical diagonalization, while SQD approximates ground-state energies through direct sampling and classical post-processing for error mitigation. SKQD integrates these concepts, inheriting the convergence guarantees and noise-tolerant characteristics of both. As a result, it is particularly well suited for near term quantum devices and for strongly correlated problems such as quantum magnetism and material simulations.

In SKQD, the effective Hamiltonian is constructed within the quantum Krylov subspace generated by time-evolved reference states. For a system described by a Hamiltonian $H$ acting on an $N = 2^n$-dimensional Hilbert space of $n$ qubits and an initial state $|\psi_0\rangle$, the Krylov basis is defined as
\begin{equation}
|\psi_k\rangle = e^{-i k H \Delta t} |\psi_0\rangle, \quad k = 0, 1, \dots, d-1,
\end{equation}
where $\Delta t$ is a discrete time step and $d$ denotes the chosen subspace dimension~\cite{yu2025sample}.
The projected effective Hamiltonian $H_{\mathrm{eff}}$ and overlap matrix $S_{\mathrm{eff}}$ are then obtained from quantum measurements of the corresponding matrix elements
\begin{equation}
(H_{\mathrm{eff}})_{ij} = \langle \psi_i | H | \psi_j \rangle, \qquad
(S_{\mathrm{eff}})_{ij} = \langle \psi_i | \psi_j \rangle,
\end{equation}
where the $| \psi_j \rangle$ consists of bitstrings of length $n$.
Diagonalization of the generalized eigenvalue problem
\begin{equation}
H_{\mathrm{eff}} v = \lambda S_{\mathrm{eff}} v
\label{eq:diagonalization}
\end{equation}
yields approximations of the low-energy eigenvalues $\lambda$ and the corresponding joint publication of the results  $v$ defining the approximate eigenstates in the Krylov basis.

The SKQD performance guarantees require the Hamiltonian, its ground state, and the Krylov evolution initial state to obey certain requirements, such as a well-behaved spectral property, non-trivial initial state overlap with the true ground state and ground state sparsity. Let us recall the sparsity definition from \cite{yu2025sample}:
\begin{definition}[($\alpha_L$, $\beta_L$)-sparsity]
For any state $\ket{\psi}$ let
\begin{equation}
    \ket{\psi}=\sum_{k}g_k\ket{b_k}
\end{equation}
with ordering $|g_1|\ge|g_2|\ge\dots$. We say that $\ket{\psi}$ exhibits $(\alpha_L, \beta_L)$-sparsity if
\begin{equation}
    \sum_{k=1}^{L}|g_k|^2\ge\alpha_L
\end{equation}
and
\begin{equation}
    |g_k|^2\ge\beta_L,\qquad 1\le k\le L.
\end{equation}
\end{definition}

\begin{figure*}
    \centering
    \includegraphics[width=0.5\linewidth]{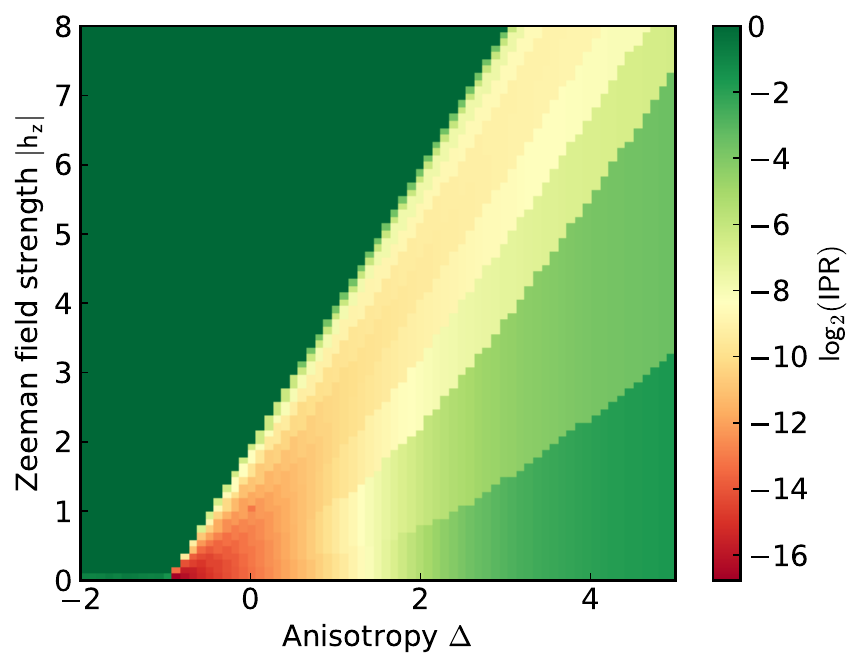}
    \caption{Ground state sparsity for 20 spin system with $J=1$ in $(\Delta, h_z)$-space plotted as the logarithm of the inverse participation ratio.}
    \label{fig:sparse}
\end{figure*}

To illustrate how the sparsity of the ground state of (\ref{eq:genericHamiltonian}) changes with respect to the anisotropy and strength of the external magnetic field, Figure \ref{fig:sparse} plots the logarithm of the Inverse Participation Ratio of the ground state in the $(\Delta, h_z)$ space for a 20 spin system. It is a discrete version of the familiar XXZ-Heisenberg Hamiltonian phase diagram that clearly illustrates the trivial ferromagnetic region and the transition from a gapped system to a gapless critical region depending on the particular choice of $(\Delta, h_z)$. The discontinuities mark individual particle/magnetization sectors of the 20 qubit systems. The color spectrum of Figure \ref{fig:sparse} thus naturally suggests the regimes where SKQD performance might struggle due to the lack of ground state sparsity.

Compared with standard variational quantum eigensolver (VQE) techniques, SKQD eliminates the need for parameter optimization, making it less vulnerable to barren-plateau effects and initialization sensitivity. Instead, its performance depends primarily on controlled time-evolution accuracy and sampling precision, both of which scale favorably on near-term quantum hardware. This combination of algorithmic simplicity, robustness to noise, and spectral accuracy positions SKQD as a promising framework for quantum simulations of correlated spin systems and Hamiltonians with dense ground-state structure.


\section{Benchmark Setup}




\subsection{Model Hamiltonians and Lattice Geometries}
We benchmark the SKQD algorithm on the spin–$\tfrac12$ Heisenberg family in both one and two spatial dimensions. 
Open boundary conditions are used
throughout.  In 1D every qubit represents one lattice site of a linear chain. For 2D lattices, an $N$-qubit register is mapped onto the \emph{most-square}
open rectangle $r\times c$ ($N = r c$ with $r\!\le\!c$) so that all qubits have
nearest-neighbor connectivity along both axes.
The rectangle is chosen by the simple divisor-search heuristic

\begin{equation}
  (r,c)=
  \mathop{\arg\!\min}_{\substack{r c = N\\r\le c}}
  \bigl|\,r-c\,\bigr|,
  \qquad    
  r=\max\bigl\{\,d\le\sqrt{N}\,\big|\,N\bmod d = 0\bigr\},
  \label{eq:A1}
\end{equation}

\noindent%
where the search starts at $\lfloor\sqrt{N}\rfloor$ and steps downward until the
first divisor $d$ is found; $c=N/d$ follows directly~\footnote{For prime $N$ the loop stops at $d=1$, and the layout degenerates to a
  $1\times N$ strip with nearest-neighbor connectivity.}.

\subsection{Initial State Preparation}
 The choice of the initial state is of critical importance for the success, precision, and efficiency of the SKQD algorithm and its predecessors. The quality of this initial state is primarily determined by its overlap with the true ground state of the quantum system being studied. The convergence requirement for SKQD is an initial state with $\tfrac{1}{\mathrm{poly}(n)}$ overlap with the ground state, together with a sparse ground state.
 The accuracy of the ground state energy approximation is directly tied to the magnitude of this overlap \( \gamma_0^{\mathrm{2}} = \left| \langle \phi_0 \mid \psi_0 \rangle \right|^{\mathrm{2}} \). The authors in \cite{doi:10.1137/21M145954X} note that, in the noiseless case, the energy error obtained from the Krylov quantum subspace approach is inversely proportional to the quality of the initial state overlap \( |\gamma_0|^2 \). Additionally, the error in SKQD is determined by the ``sparsity'' of the ground state, and only \textit{indirectly} tied to the overlap. In the main theorem from \cite{yu2025sample}, it is stated that the error in the ground state energy $\Delta E_0$ is bounded by\[ \Delta E_0\leq \sqrt{8} \, \|H\| \, \left( 1 - \sqrt{\alpha _L^{(0)}} \right)^{1/2} . \]
Note that initial state overlap \( |\gamma_0|^2 \) is not explicitly mentioned in this final error formula. In fact, the accuracy is only guaranteed if we successfully sample all the important components (``bitstrings'') of the ground state. The practicality of this is governed by the number of samples M required from each Krylov basis state. Theorem 1 in  \cite{yu2025sample}, specifies that M must exceed \[ M > \frac{d^2 \log(L / \eta)}{|\gamma_0|^2 \, \left( \beta_L^{(0)} - 2\sqrt{\tilde{\varepsilon}} \right)} \] 
where the number of samples required scales inversely with the overlap of the initial state \( |\gamma_0|^2 \). 
In summary, for the SKQD method, a poor choice of initial state does not necessarily worsen the theoretical best-case error bound, but it can make the quantum measurement cost required to reach that bound, prohibitively large.

\subsubsection{Singlet State}
The single product state provides a physically-motivated, high-quality first guess, that is known to be adiabatically connected to the true ground state.
To introduce the construction more clearly, we begin with the formula and corresponding circuit for a two-qubit singlet system
\[
|\Psi^-\rangle = \frac{1}{\sqrt{2}} (|01\rangle - |10\rangle) = U_{\text{singlet}} |00\rangle
\]
\begin{center}
    \begin{quantikz}
    \lstick{\(|0\rangle\)} & \gate{X} & \gate{H} & \ctrl{1} & \qw \\
    \lstick{\(|0\rangle\)} & \gate{X} & \qw       & \targ    & \qw
    \end{quantikz}
\end{center}

In \cite{PhysRevResearch.5.013183} the authors identify the singlet state as a suitable starting-point ansatz  for \(\Delta\) in the AFM range, arguing that the AFM Heisenberg model favors anti-aligned spins, and a singlet pair \mbox{$\left(|01\rangle - |10\rangle\right)/\sqrt{2}$} is the quintessential quantum state of two perfectly anti-aligned spins. Although we do not make use of their adiabatic connection, we adopt the same physical reasoning to justify its use in our SKQD simulations.

For many of our numerical experiments, we choose the singlet product state as the initial state
\[
\ket{\psi_{\text{singlets}}} = \frac{1}{\sqrt{2^{N/2}}} \prod_{j=1}^{N/2} (\ket{01} - \ket{10})_{2j-1, 2j}
\]
This state is the ground state of the Hamiltonian with interaction only on odd bonds, for even N
\[
\hat{H}_{\text{odd}} = \sum_{j=1}^{N/2-1} \left( \sigma_x^{[2j-1]} \sigma_x^{[2j]} + \sigma_y^{[2j-1]} \sigma_y^{[2j]} + \Delta \sigma_z^{[2j-1]} \sigma_z^{[2j]} \right),
\]
making it a natural and physically well-motivated choice for our problem setting.

\subsubsection{Single Particle Sector Sweep}
\label{subsection:wstates}

One disadvantage of the singlet initial state presented above and other half-filling states (like N\'eel state) is that their half-filling Hamming weight is being propagated along the Krylov evolution, hence all the bitstrings sampled (in a hypothetical noiseless setting) are going to be half-filled as well. While this is working for example in isotropic AFM Hamiltonian setting ($J=1$, $\Delta = 1$), in case we  decide to break the symmetry with Zeeman field, half-filled initial states are no more suitable candidates to construct a ground state approximation. As a consequence, the quality of the approximated ground state energy using singlets as the initial state will degrade as the system's ground state gets more polarized (i.e. shifted away from half-filling). We need to be able to sample bitstrings from different particle sectors as well to make sure we receive a good approximation of the system's ground state. Taking into account that our Zeeman field will be pointing in one direction only, we need to sweep through all particle sectors from the zero state to half-filling states.

For a particle sector with $k$ particles, we divide the qubits into $k$ groups and initialize the state as a product of $W$-state, one over each group:
\begin{equation}
\label{eq:init_wstates}
\ket{\psi_0^k}:=\bigotimes_{m=0}^{k-1}W_{\lfloor(m+1)N/k\rceil-\lfloor mN/k\rceil},
\end{equation}

where $W_{m}$ is an $m$ qubit $W$-state and $\lfloor x\rceil$ is a rounding operation to the nearest integer. Being a product of $W$-states the preparation of $\ket{\psi_0^k}$ can be done efficiently with $O(N/k)$ circuit depth.

It is apparent that $\ket{\psi_0^k}$ contains only $k$-particle bitstrings. Moreover, the choice of $W$-states ensures that the particles are spread evenly as, for example, $\ket{\dots 11\dots}$ bitstring is not present, favoring the AFM particle ordering.

In order to approximate the ground state magnetization, we first construct an initial state (\ref{eq:init_wstates}) lying in the particular particle sector. Then we calculate the ground state energy approximation using SKQD. Since the particle sector is preserved through the evolution, we get (in the noiseless setup) a ground state energy approximation restricted to a single particle sector. To get the real ground state approximation, we do the sweep for all particle sectors from zero particles (trivial) to half-filling, and whichever has the smallest energy, we designate it to be the ground state approximation of the system. The value of magnetization is then equal to the magnetization characteristic for the particle sector, which delivered the lowest energy in the sweep. To save computational time, we choose for each value of the Zeeman term only particle sectors close to the value of the reference DMRG calculation, as particle sectors far away from the true ground state are unlikely to produce even lower energy approximation. Please refer to Appendix \ref{app:mag_filtering} for a detailed description.



\begin{figure}[t]
\includegraphics[width=0.9\linewidth]{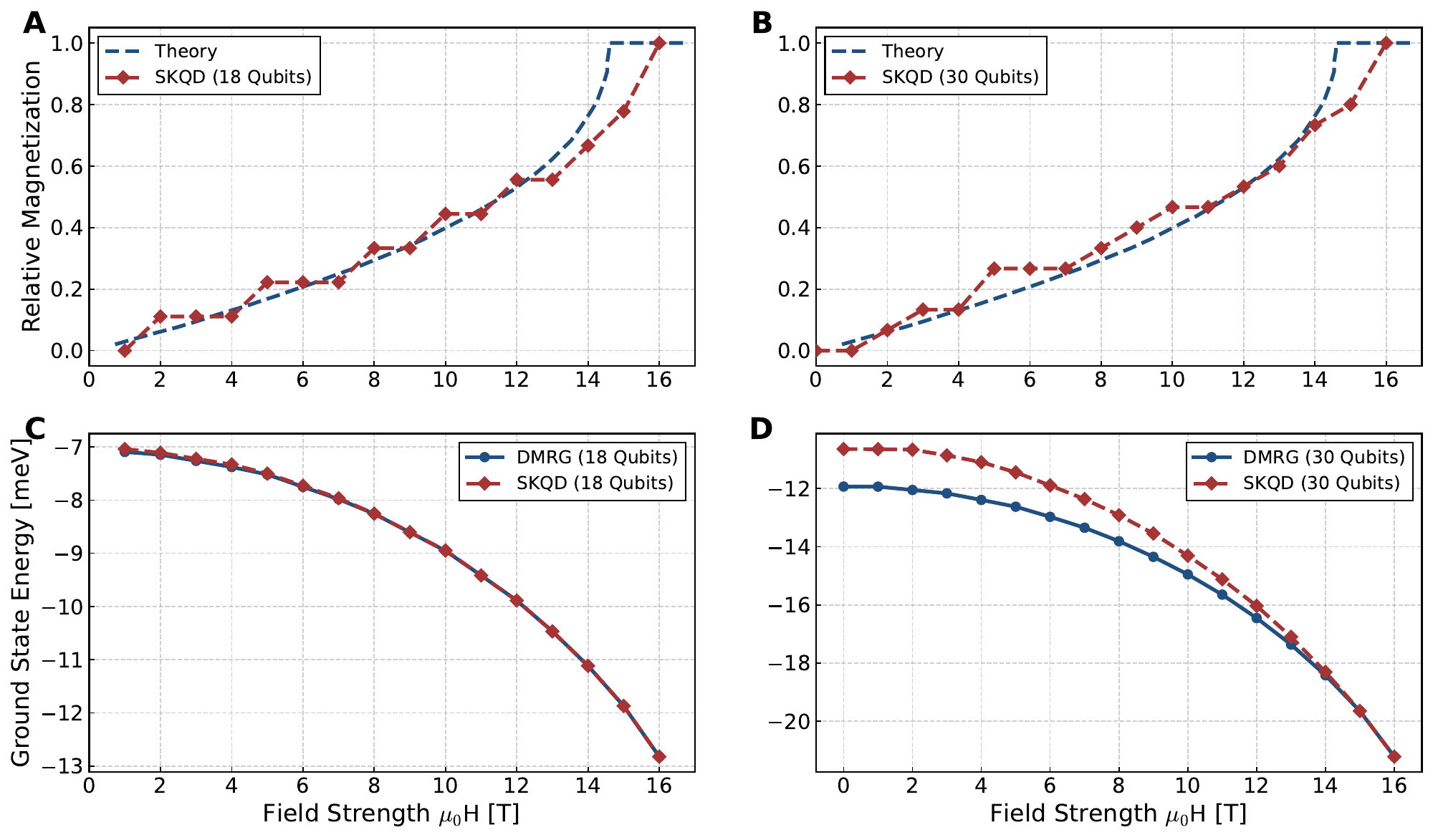}
\caption{Relative Magnetization (\textbf{A}, \textbf{B}) and ground state energy (\textbf{C}, \textbf{D}) of CuPzN approximated with SKQD for 18 (\textbf{A}, \textbf{C}) and 30 (\textbf{B}, \textbf{D}) spin systems. Compared with theoretical values derived from Bethe ansatz \cite{PhysRevB.59.1008} and state-of-the art DMRG calculations.}
\label{fig:results_field_sweep}
\end{figure}
 \section{Results}
 \label{section:results}
The performance of SKQD across the antiferromagnetic and gapless regimes, on one- and two-dimensional Heisenberg models is investigated using a combination of quantum hardware experiments, exact and DMRG benchmarks, and simulations on small 2D lattices. 
Our results show that SKQD reliably captures ground-state energies and magnetization in the antiferromagnetic phase, where the sparsity is well conditioned. 
Moreover, even in regimes where the ground state exhibits a dense structure and sparsity is absent, SKQD reproduces the correct qualitative trends.
For the underlying Hamiltonian, we chose the quasi-1D model of Copper Pyrazine Dinitrate (CuPzN) as presented in \cite{PhysRevB.59.1008} resulting in isotropic (XXX) Hamiltonian setting with the coupling constant $J=0.91\,\mathrm{meV}$. 
The other parameters used to setup and run the SKQD simulations on quantum hardware are reported in Appendix \ref{app:setup}.

Figures \ref{fig:results_field_sweep} (\textbf{A} and \textbf{B}) show the magnetization of CuPzN reconstructed with SKQD on 18- and 30-qubit IBM Heron quantum processor. 
Across the entire field range, SKQD closely follows the Bethe ansatz reference curve, including the nonlinear rise in the intermediate-field region and saturation near the critical field. 
This agreement is notable because the low-field isotropic regime corresponds to the hardest region in the sparsity diagram, see Figure~\ref{fig:sparse}, where the ground state is highly delocalized.

The energy estimates in Figures \ref{fig:results_field_sweep} (\textbf{C} and \textbf{D}) further validate the performance of SKQD. 
For 18 qubits, SKQD tracks DMRG with excellent accuracy. 
For 30 qubits, deviations appear in the weak-field regime, consistent with reduced sparsity and increased sampling demands, but systematically diminish as the external field polarizes the ground state. 
As seen in Figure~\ref{fig:sparse}, weaker fields place the system in a challenging regime for SKQD, $(\Delta, h_z)\sim (1, 0)$, which implies increasing classical postprocessing demands as the system size grows.
The inverse participation ratio map in Figure~\ref{fig:sparse} provides a unified explanation of SKQD’s behavior. 
Accuracy is reduced in low-field, isotropic regions where the ground state is highly entangled and non-sparse, but improves distinctly
as either field strength or anisotropy increases. 
This trend appears consistently in both magnetization and energy benchmarks and matches the expected dependence of SKQD on the number of significant basis components required in the Krylov reconstruction.

To extract magnetization on hardware, we employ a magnetization-sector sweep using W-state initializations tailored to each conserved particle number. 
Post-selection on excitation number (Appendix B) is critical: although incorrect bitstrings do not affect the minimum-energy estimate, they bias magnetization. Filtering restores correct sector ordering and yields smooth and physically accurate curves.

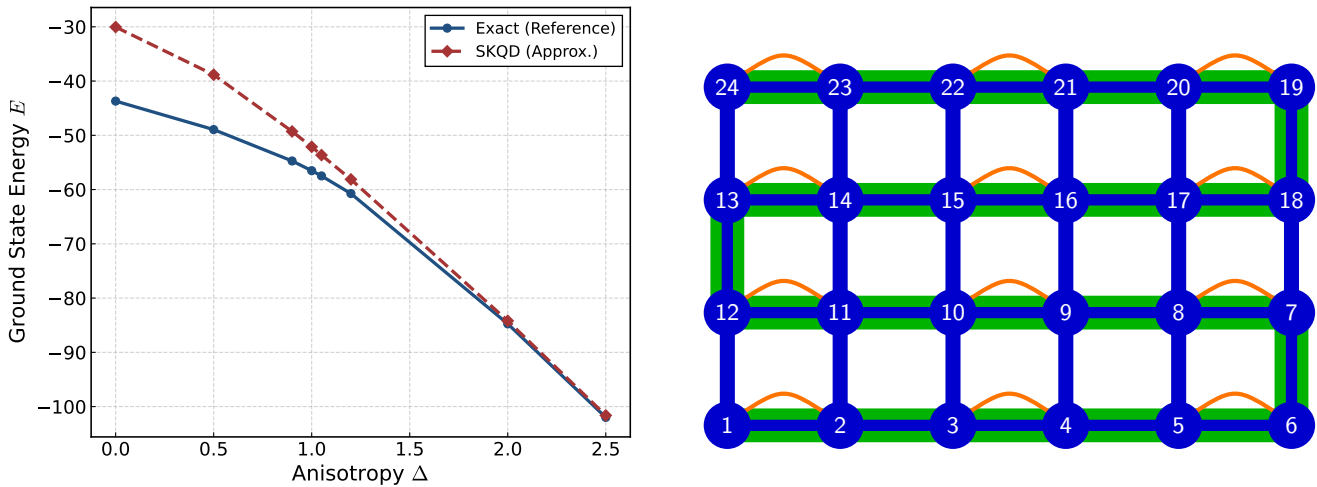
\begin{figure}[t]
    \RawFloats
    \centering
    \begin{minipage}{0.99\textwidth}
    \begin{minipage}{0.48\textwidth}
        \centering 
        \includegraphics[width=\linewidth]{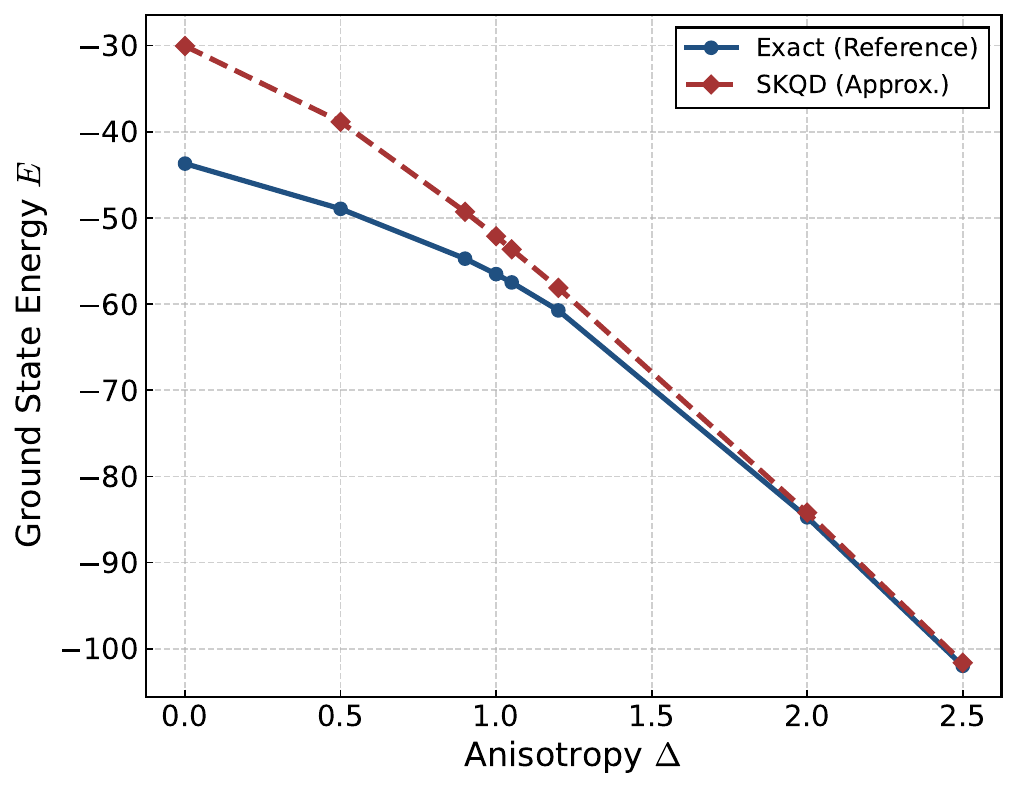}
    \end{minipage}
    \hfill
    \begin{minipage}{0.48\textwidth}
        \centering 
        \begin{tikzpicture}[
    scale=1.5,
    site/.style={circle, draw=blue!80!black, fill=blue!80!black, thick, minimum size=6mm, inner sep=0pt},
    connection/.style={blue!80!black, line width = 2mm},
    snake/.style={green!70!black, line width = 1.5mm},
    coupling/.style={orange!90!red, ultra thick, line cap=round}
]

    \def\cols{6}
    \def\rows{4}

    \foreach \y in {0,...,\numexpr\rows-1} {
        \foreach \x in {0,...,\numexpr\cols-1} {
            \ifnum\x<\numexpr\cols-1\relax
                \draw[connection] (\x,\y) -- (\x+1,\y);
            \fi
            \ifnum\y<\numexpr\rows-1\relax
                \draw[connection] (\x,\y) -- (\x,\y+1);
            \fi
        }
    }

    \draw[snake] (0,0+0.1) 
    \foreach \y in {0,...,\numexpr\rows-1} {
        \ifodd\y
            -- (0, \y+0.1)
            \ifnum\y<\numexpr\rows-1\relax
                -- ++(0.1, 0) -- ++(0, 1) -- ++(0.0, 0)
            \fi
        \else
            -- (\cols-1, \y+0.1)
            \ifnum\y<\numexpr\rows-1\relax
                -- ++(0.1, 0) -- ++(0, 1) -- ++(-0.0, 0)
            \fi
        \fi
    };

    \draw[snake] (0,0-0.1) 
    \foreach \y in {0,...,\numexpr\rows-1} {
        \ifodd\y
            -- (0, \y-0.1)
            \ifnum\y<\numexpr\rows-1\relax
                -- ++(-0.1, 0) -- ++(0, 1) -- ++(0.0, 0)
            \fi
        \else
            -- (\cols-1, \y-0.1)
            \ifnum\y<\numexpr\rows-1\relax
                -- ++(-0.1, 0) -- ++(0, 1) -- ++(-0.0, 0)
            \fi
        \fi
    };

    \foreach \y in {0,...,\numexpr\rows-1} {
        \foreach \x in {0, 2, ..., \numexpr\cols-2} {
            \draw[coupling] 
                ([yshift=2pt, xshift=2pt] \x, \y) .. controls ([yshift=10pt, xshift=0pt] \x+0.5, \y) .. ([yshift=2pt, xshift=-2pt] \x+1, \y);
        }
    }

    \foreach \y in {0,...,\numexpr\rows-1} {
        \foreach \x in {0,...,\numexpr\cols-1} {
            \pgfmathparse{int(\y*\cols + (mod(\y,2) == 0 ? \x : \cols-1-\x) + 1)}
            \let\siteindex\pgfmathresult
            \node[site] at (\x,\y) {\small \textcolor{white}{\textsf{\siteindex}}};
        }
    }
\label{tikz:snake-like}
\end{tikzpicture}
    \end{minipage}
    \caption{Comparison between SKQD energy estimates and exact energies on a 6x4 square lattice (left) and snake-like initial state (green) on a 2D square lattice layout used for SKQD simulation (right). The red lines show the state's singlet couples.}
    \label{fig:snake_layout}
    \vfill
    \end{minipage}
\end{figure}
Motivated by the forthcoming availability of IBM Nighthawk Quantum processor, which features four-nearest-neighbor square lattice connectivity, and to assess performance of beyond 1D, we apply SKQD to a 6×4 square lattice using a shallow snake-like singlet initialization, see Figure \ref{fig:snake_layout}. 
Despite the increased connectivity of the lattice and the dense ground-state structure, SKQD reproduces the overall monotonic dependence of the ground-state energy on anisotropy and captures the correct qualitative trends when compared with exact diagonalization across much of the
 $\Delta$ range. 
 Agreement improves toward the Ising limit, consistent with the sparsity trends observed in one dimension. 
 These results indicate that SKQD can be applied to higher-dimensional lattice geometries while retaining qualitatively reliable behavior.

 
\section{Conclusions and Outlook}
The performance and applicability of the SKQD method on the Heisenberg family of spin–$\frac12$ models in both one and two spatial dimensions was evaluated in this work. 
Despite the strongly correlated nature of these models and the resulting dense ground-state structure, well outside the sparsity assumptions underlying the formal efficiency analysis of SKQD, we find that the method exhibits consistent convergence and captures qualitative trends across regimes ranging from easy-plane anisotropy to the isotropic point and deep into the Ising-like region.
This observation highlights the practical applicability of SKQD beyond the sparsity regime assumed in existing analyses.
We benchmark SKQD against the Bethe ansatz and Density Matrix Renormalization Group (DMRG) calculations. 
While numerical discrepancies stemming from finite sampling, approximate Krylov-space construction, and batch truncation naturally remain, the comparison with ED and DMRG demonstrates that SKQD captures the correct physical behavior and that its performance improves systematically as anisotropy increases and effective sparsity becomes more favorable.
These benchmarks provide a physically meaningful validation of the algorithm and highlight the regions of the phase diagram where SKQD is most effective and confirm that SKQD remains robust even when sparsity assumptions are not strictly satisfied.

Beyond energy estimation, we explored how initialization strategies influence convergence and observable accuracy.
Physically motivated initial states such as singlet products yield strong performance in the antiferromagnetic regime, while W-state sector sweeps enable accurate magnetization extraction by exploiting underlying U(1) symmetry. 
Our results show that SKQD can reliably identify magnetization sectors and reproduce field-dependent spin polarization trends, providing evidence that the method can access more complex observables than just ground-state energies.
Our preliminary 2D simulations further illustrate that SKQD extends naturally to higher-dimensional lattice geometries.
We demonstrate quantum hardware results for 18 and 30 qubit Heisenberg chains. These simulations provide a direct comparison between SKQD on real quantum devices and DMRG baselines, enabling the first hardware demonstration of SKQD applied to magnetization curves and field-driven transitions.

Overall, our study demonstrates that SKQD is a flexible quantum-classical method for correlated spin systems, capable of scaling for appropriate ground state sparsity regimes and achieving meaningful agreement with state of the art classical methods such as DMRG while remaining implementable on near-term quantum hardware. With the forthcoming magnetization experiments and extended 2D benchmarks, SKQD stands as a promising tool for quantum simulation in both algorithmic and materials-science contexts.

\section*{Acknowledgments}

We gratefully acknowledge insightful discussions and continuous support from Axel Sauerland, Gabriele Compostella, Arne-Christian Voigt, Rukhsan Ul Haq, Ben Jaderberg, Will Kirby, Mirko Amico, Kunal Sharma, Antonio Mezzacapo, Keerthi Kumaran, Jennifer Glick, and the IBM Quantum Support Team throughout the development of this work. Their feedback on both the physics and the implementation of the SKQD benchmarks has been invaluable.



\newpage
\appendix
\section{Exploring Magnetization Sectors with W States}
In a similar fashion to the KQD approach, the SKQD algorithm combined with the use of W states as initial states allows us to predict the correct magnetization in the presence of a longitudinal field $h_z$ by explicitly exploiting the underlying $\mathrm{U}(1)$ symmetry. We perform a magnetization-sector sweep, considering each sector $k$ of excited particles separately. Each such sector corresponds to a magnetization
\[
M = N - 2k,
\]
which follows from the definitions $Z\ket{0} = \ket{0}$ and $Z\ket{1} = -\ket{1}$, with the total magnetization operator defined as
\[
M = \sum_{j} Z_j,
\]
and the initial states constructed as described in Sec.~\ref{subsection:wstates}. For the sector-$k$ W state $\ket{W_{N,k}}$, we therefore have
\[
\braket{W_{N,k} | M | W_{N,k}} = N - 2k.
\]
We find perfect agreement between the benchmarked magnetization and the magnetization obtained by selecting the sector whose approximate ground-state energy is lowest. This procedure is directly aligned with the ideas of the KQD method, where one works within fixed-$k$ (particle-number or magnetization) symmetry sectors and chooses an initial state residing in that sector before performing the Krylov-space diagonalization.

\begin{figure}[ht]
    \centering
    \includegraphics[width=0.5\linewidth]{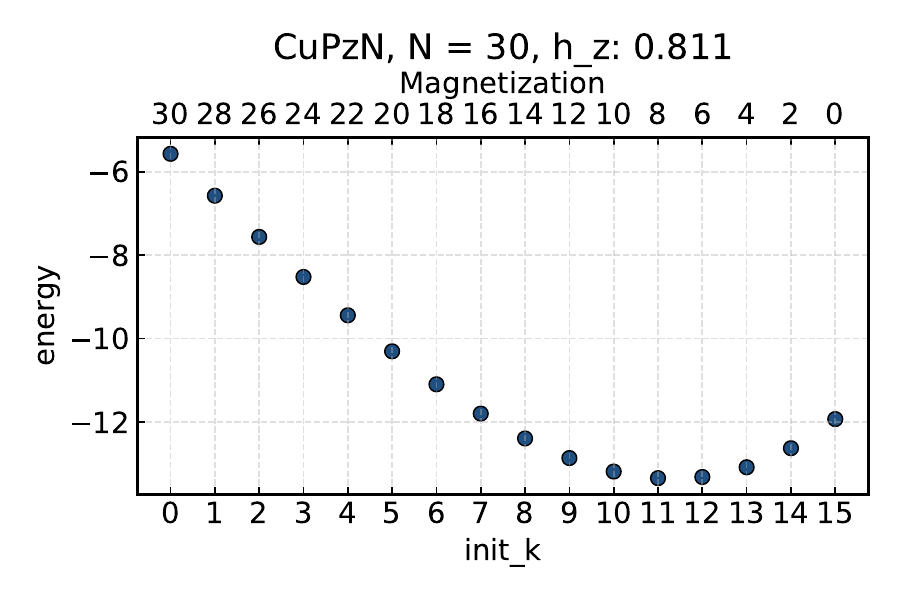}
    \caption{Magnetization Sector in the presence of external field.}
    \label{fig:magnetization_sector}
\end{figure}

\section{Postprocessing: Magnetization Sector Filtering}
\label{app:mag_filtering}

During the field-sweep hardware runs, we initially observed that several magnetization sectors reported identical (and correct) ground state energy values, producing a flat behavior, unlike what is seen in Fig.~\ref{fig:magnetization_sector}. While this is sufficient for estimating the ground-state energy, it prevents a reliable determination of the magnetization. 

This effect arises from measurement errors: incorrect bitstrings do not alter the estimated minimum eigenvalue of the matrix in Eq.~\eqref{eq:diagonalization}, provided that all relevant correct bitstrings occur. However, such errors are detrimental for estimating the magnetization, which depends on the population of bitstrings rather than just the lowest energy.

\begin{figure}[ht]
    \centering
    \includegraphics[width=0.5\linewidth]{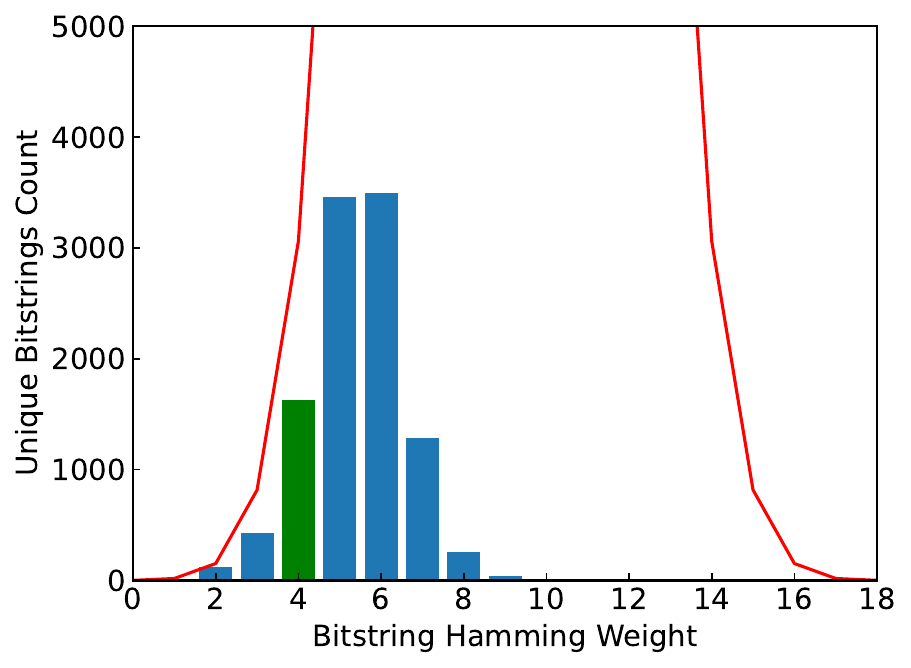}
    \caption{Histogram of sampled unique bitstrings on 18 spin system. Green bar marks the initialization particle sector and red line shows the maximum number of unique bitstrings, explaining the sampling distribution mean value shift towards the half-filling sector in the middle.}
    \label{fig:spillover}
\end{figure}

Quantum hardware errors can occur anywhere in the circuit execution pipeline (gate imperfections, decoherence, readout noise, etc.). In the bitstring distribution shown in Figure \ref{fig:spillover}, we observe that for a fixed-$k$ evolution, where the model should conserve total excitation, some measured bitstrings nevertheless contain too few or too many excitations.

To address this, we discard all bitstrings whose excitation number does not match the initial one. Although this filtering step is wasteful, we found that it yields a reliable identification of the magnetization sector: only the physically correct sector exhibits the lowest postselected energy.

\section{Setup and Parametrization of Calculations}
\label{app:setup}

Throughout all simulations we fix the time step to $\mathrm{d}t = 0.3$ and use the same initial state as described above. 
To mitigate the non-sparsity of the ground state and reliably capture all relevant bitstrings, we employ  
$300{,}000$ shots per Krylov dimension for the $18$-qubit system and $600{,}000$ shots for the $30$-qubit system, 
with a total of five Krylov steps. 
At each step, the time-evolution operator is approximated by a second-order Trotter circuit with three repetitions.  
The results of the field sweep are summarized in Fig.~\ref{fig:results_field_sweep}.

To obtain essentially exact reference data for the ground-state energy and magnetization, we perform classical simulations using the density-matrix renormalization group (DMRG) in a matrix-product-state (MPS) formulation. We consider a spin-$\tfrac{1}{2}$ chain of length $L$ with open boundary conditions and couplings $J_x$, $J_y$, $J_z$ and a longitudinal field $h_z$, as implemented in the models of the TeNPy library~\cite{tenpy2024}. Exploiting the underlying $\mathrm{U}(1)$ symmetry, we work in the $S^z$-conserving sector, and for each value of the longitudinal field we perform a sweep over total-magnetization sectors $q$, initializing the MPS from simple product states with the corresponding numbers of up and down spins. For every sector we run finite-system DMRG with a maximum bond dimension $\chi_{\max} = 128$, up to four sweeps, and a truncation threshold for singular values of $\mathrm{svd}_{\min} = 10^{-10}$. We then select the sector yielding the lowest variational energy as the ground state for that field value. From the converged MPS we extract both the ground-state energy and the corresponding total magnetization $\langle S^z_{\mathrm{tot}} \rangle$, which we use as classical benchmark data for the quantum algorithms studied in this work.

All quantum hardware results in this work are run on \textit{}{ibm\_marrakesh} 156-qubits IBM Heron r2 superconducting qubits QPU. The total hardware time was 7,5 minutes resp. 15 minutes for 18 resp. 30 qubits single particle sector SKQD calculation with the parameter setup mentioned above.

\bibliography{ref}

\end{document}